\documentclass[twocolumn,secnumarabic,amssymb, nobibnotes, aps, prd,superscriptaddress]{revtex4-2}

\usepackage{graphicx}% Include figure files
\usepackage{dcolumn}% Align table columns on decimal point
\usepackage{bm}% bold math
\usepackage{tabularx}
\usepackage{upgreek}
\usepackage{color}
\usepackage{xcolor}

\usepackage{ulem}

\begin{document}

\title{Molecular dynamics insights into the Debye process of 1-propanol}

\author{Marceau H\'enot}
\email{marceau.henot@cea.fr}
\affiliation{SPEC, CEA, CNRS, Université Paris-Saclay, CEA Saclay Bat 772, 91191 Gif-sur-Yvette Cedex, France.}

\author{Jan Philipp Gabriel}
\email{Jan.Gabriel@dlr.de}
\affiliation{Institute of Frontier Materials on Earth and in Space, German Aerospace Center, 51170 Cologne, Germany}

\date{\today}

\begin{abstract}
We reproduce the Debye process in the dielectric response of liquid 1-propanol by all-atom molecular dynamics simulations between 340 K and 200 K. The analysis of dipolar correlations reveals that the $\alpha$ relaxation originates from hydrogen-bond (HB) breaking, while the dominant Debye process results from long-ranged cross-correlations extending over several molecular diameters. By separating intra- and extra-cluster contributions, we demonstrate that HB supramolecular clusters account for the main part of the static dielectric response, with extra-cluster molecules playing only a minor role at low temperatures. Besides, clusters do not preserve connectivity on timescales exceeding the $\alpha$ relaxation time. This indicates that clusters stabilize orientational correlations beyond individual molecular relaxation times by transmitting alignment to newly incorporated molecules. These findings provide microscopic evidence that the Debye relaxation in mono-alcohols arises from the collective dynamics of HB clusters and offer a framework to study dielectric spectra in other hydrogen-bonded liquids.
\end{abstract} 

\maketitle

Dielectric spectroscopy (DS) has, for more than a century, provided key insights into molecular dynamics by probing the reorientation of molecules over a broad frequency range~\cite{kremer2002broadband}. Because polar molecules tend to exhibit short-range orientational correlations~\cite{Kirkwood1939, bottcher1973, dejardin2022temperature}, DS is sensitive not only to the dynamics of individual molecules but to dipolar cross-correlations as well~\cite{bottcher1978}. This can affect the shape of their relaxation spectra~\cite{bohmer2024dipolar}, as recently highlighted through comparisons with light-scattering experiments~\cite{gabriel2017debye, pabst2021generic, paluch2023determination, bohmer2025spectral}. A particularly striking case is found in mono-alcohols~\cite{bohmer2014structure}, which display an intense and unusually narrow peak known as the Debye relaxation~\cite{debye_polar_1929, hansen1997dynamics, kudlik1997slow}. This process leaves little to no signature in other experimental techniques probing the faster structural $\alpha$ relaxation~\cite{gainaru2010nuclear, hecksher2016communication, gabriel2017debye, weigl2019local}. While the role of supramolecular hydrogen-bonded (HB) structures has long been acknowledged~\cite{levin1982dipole, kalinovskaya2000exponential, singh2012watching}, the microscopic mechanism underlying the Debye relaxation remains under debate. In the transient chain model (TCM)~\cite{gainaru2010nuclear}, it is attributed to the reorientation of the end-to-end dipole of HB chains through the attachment or detachment of units at their ends. In the \textit{diffusive La Ola wave} (DLO) picture~\cite{wieth2014dynamical}, chains polarize surrounding molecules and transmit an excitation over larger scales and longer timescales. More recently, an analogy was suggested with the living polymer model (LPM) in which the Debye process results from a balance between chain diffusion and breaking~\cite{patil2023molecular, cheng2024hydrogen}. Molecular dynamics (MD) simulations could help advance this question, as they have been useful in better understanding a variety of molecular relaxation processes~\cite{laage2006molecular, laage2008molecular, wieth2014dynamical, atawa2019molecular, guiselin2022microscopic, baptista2023chilling, pabst2025glassy} and lately in disentangling self and cross dipolar correlations~\cite{koperwas2022computational, alvarez2023debye, henot2023orientational, pabst2025salt, henot2025emergence, henot2025, koperwas2025can}.

In this letter, we study the dielectric response of 1-propanol using MD simulations over a fairly wide temperature range, going from 340 to 200~K, for which we demonstrate that relaxation spectra comparable to experiments can be obtained. We disentangle the underlying self and cross-correlations, and we investigate the length scale over which this latter contribution builds up. We then characterize supramolecular HB structures from snapshots of the simulation. This allows us to study the relative contributions to the evolution of static dielectric permittivity of molecules involved or not in HB structures. Finally, we investigate the role played by these supramolecular structures in the Debye peak appearing in dielectric spectra. 

%%%%%%%%%%%%%%%%%%%%%%%%%%%%%%%%%%%%%%%%
\begin{figure}[htbp]
 \centering
 \includegraphics[width=\linewidth]{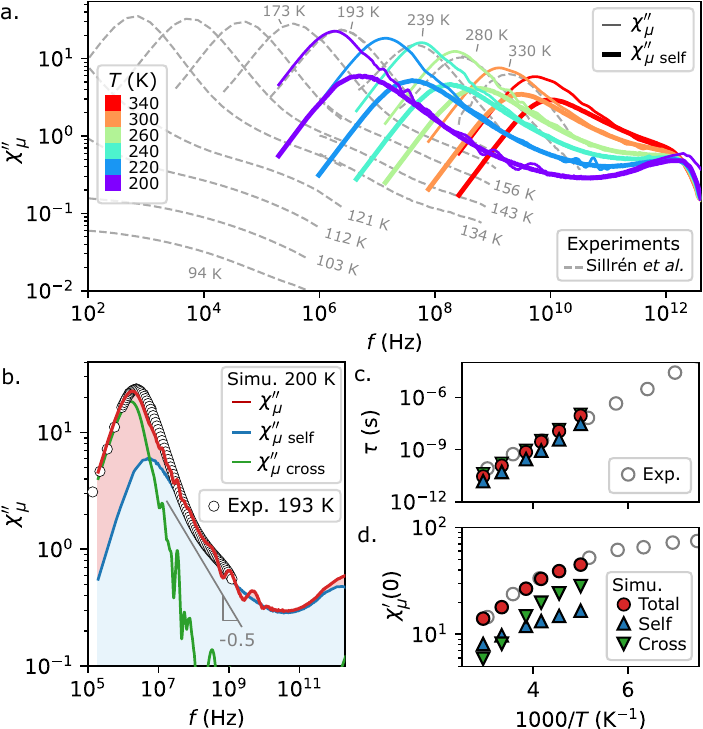}
 \caption{(a) Dielectric loss spectra obtained from simulations (total and self, shown as colored lines) compared with experimental results~\cite{sillren2014liquid} (dashed grey lines). (b) Focus on the comparison between simulations and experiments near 200 K. (c) Relaxation time and (d) static permittivity plotted as functions of the inverse temperature, with experimental data~\cite{sillren2014liquid} shown in grey.}
 \label{fig1}
\end{figure}
%%%%%%%%%%%%%%%%%%%%%%%%%%%%%%%%%%%%%%%%
We performed MD simulations on a system of 9600 1-propanol molecules using the OPLSAA-3SSPP force field~\cite{alva2022improving} (see Appendix A for details on the methods).
The dielectric response of the system was computed using the virtual cavity method described in Appendix B, and in more detail in ref.~\cite{henot2025}. It has the advantage of requiring a significantly shorter total simulation time to obtain good statistics on dielectric spectra compared to the standard method involving the total dipole of the simulation box~\cite{neumann1984consistent}. The crucial advantage here is that it allows us to reach a wide temperature range. The dipole correlation function of a virtual cavity of radius $r_\mathrm{c}$ was computed as follows: 
\begin{equation}
 C_{\mu}(t) = \left\langle\vec{\mu}_i(0)\cdot\sum_{r_{ij}<r_c} \vec{\mu}_j(t) \right\rangle
 \label{eq_def_Cmu}
\end{equation}
With $\vec{\mu}_i(t)$ the dipolar moment of molecule $i$ at time $t$. The choice of the radius $r_\mathrm{c}=20$~\AA ~results from a trade-off: it has to be small enough compared to the simulation box size to be insensitive to the electrostatic boundary conditions used in the simulation~\cite{elton2014polar,henot2025}, but large enough to include all the relevant short-range cross-correlations. Dielectric spectra $\chi_\mu^{\prime\prime}(\omega)$ obtained from $C_\mu(t)$ (see Appendix B) are shown in fig.~\ref{fig1}a for each temperature and in fig.~\ref{fig1}b at 200~K. The relaxation time $\tau_\mathrm{tot} = \int_0^\infty C_\mu(t)dt$ and the static permittivity $\chi_\mu^{\prime}(0)$ are indicated by red markers in fig.~\ref{fig1}c and d, respectively. Both quantities can be compared to experimental data (shown with grey markers). The force field was optimized to reproduce the experimental static dielectric permittivity and molecular diffusion coefficient at 300~K~\cite{alva2022improving}, and it also exhibits the same temperature dependence for both quantities down to 200 K. Moreover, despite the non-polarizable nature of the force field, MD spectra show decent agreement with experimental dielectric spectra~\cite{sillren2014liquid} (shown in grey in Fig. 1a–b), particularly around the Debye relaxation peak. 

%%%%%%%%%%%%%%%%%%%%%%%%%%%%%%%%%%%%%%%%
\begin{figure*}[htbp]
 \centering
 \includegraphics[width=\linewidth]{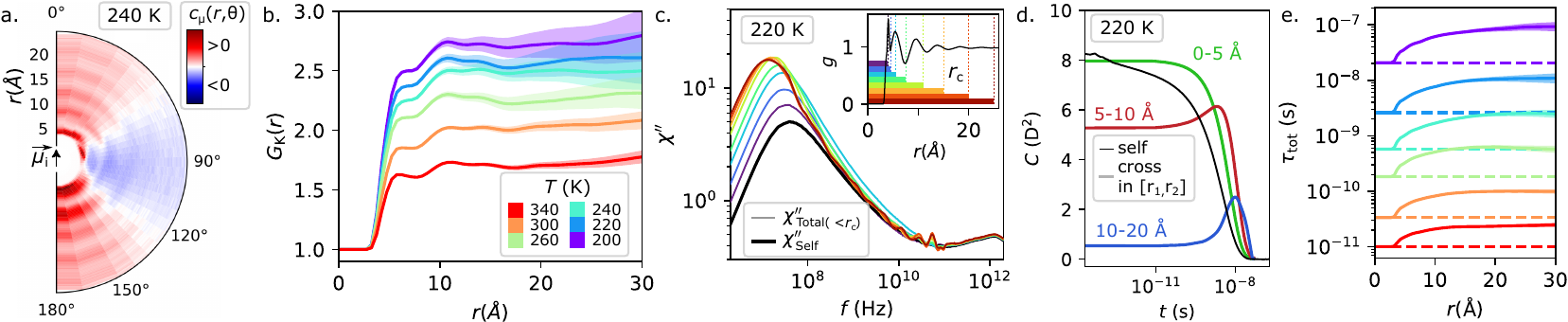}
 \caption{(a) Relative orientation and distance dependence of static cross-correlations at 240~K. (b) $r$-dependent Kirkwood correlation factor. The shading around the curves corresponds to the uncertainty originating from the fluctuations of the dipolar cross-correlations~\cite{henot2025}. (c) Relaxation spectra at 220~K when considering virtual cavities of increasing radius $r_\mathrm{c}$. The inset shows the pair distribution function of the center of charge. (d) Dipole correlation function at $220$~K for the self (black) and the cross part from shells of increasing radius (colors). (e) Total relaxation time as a function of the cavity size for each temperature.}
 \label{fig2}
\end{figure*}
%%%%%%%%%%%%%%%%%%%%%%%%%%%%%%%%%%%%%%%%
To study the effect of dipolar cross-correlations on the dielectric response, $C_\mu(t)$ can be split into self ($i=j$) and cross terms ($i\neq$j). When the dielectric permittivity is large enough, as is the case here, it is possible to similarly decompose the dielectric spectra into $\chi_\mu^{\prime\prime}(\omega)=\chi^{\prime\prime}_\mathrm{\mu~self}(\omega) + \chi^{\prime\prime}_\mathrm{\mu~cross}(\omega)$~\cite{henot2025}. In the self spectra, shown in fig.~\ref{fig1}a-b, an $\alpha$ peak is well visible, with a relaxation time $\tau_\mathrm{self} = \int_0^\infty C_\mathrm{\mu,~self}(t)dt$ that roughly follows the temperature evolution of $\tau_\mathrm{tot}$ (see fig.~\ref{fig1}c). The separation between $\alpha$ and Debye processes only starts at these high temperatures and is much smaller (a factor 3 at 340~K and 5 at 200~K) than the two decades observed at the glass transition temperature~\cite{sillren2014liquid, mikkelsen2022dielectric}. At 200~K (see fig.~\ref{fig1}b), and in contrast to the Debye peak, the $\alpha$ process appears non-exponential with a slope for the high-frequency flank close to -0.5, reminiscent of the so-called \textit{generic shape} observed experimentally for the self-relaxation of a variety of systems at deeply supercooled temperatures~\cite{pabst2021generic, bohmer2025spectral}. It should be noted, however, that we are still at a much higher temperature where more diversity is observed in the shape of the self-relaxation spectra~\cite{schmidtke2014relaxation, bohmer2025spectral}. As visible in fig.~\ref{fig1}d, the amplitude of the self contribution, shown in blue, is only weakly temperature dependent (proportional to $1/T$). In contrast, static cross-correlations, shown in green, are multiplied by 2.7 between 340~K and 200~K where they make a significant contribution to the dielectric spectrum. Interestingly, the cross part $\chi^{\prime\prime}_\mathrm{\mu~cross}(\omega)$ taken alone (see fig.~\ref{fig1}b) leads to an unusual spectral shape with a steep high frequency flank, and the experimentally observed Debye peak is only recovered when summed with the self part. This illustrates how entangled self and cross dynamics are at these high temperatures.

%%%%%%%%%%%%%%%%%%%%%%%%%%%%%%%%%%%%%%%%
\begin{figure}[htbp]
 \centering
 \includegraphics[width=\linewidth]{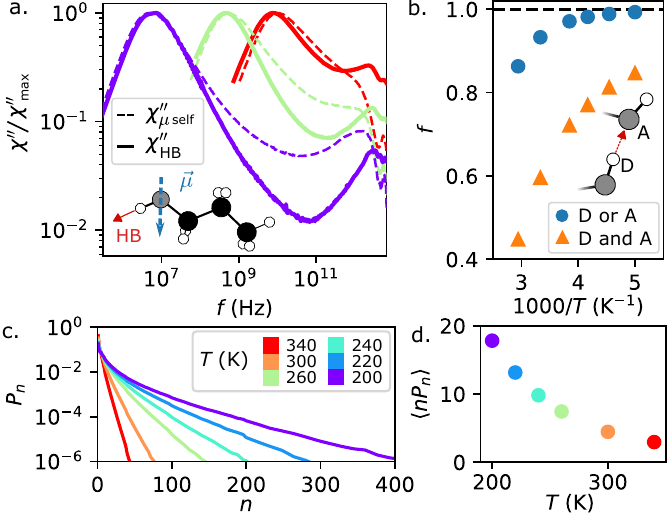}
 \caption{(a) Relaxation spectra corresponding to the self dipole orientational correlation function (dashed lines) and to the HB correlation function (solid lines), at 200, 260, and 340~K. (b) Fraction of molecule involved in a HB as a donor (D) or acceptor (A) (blue circles) or as A and D (orange triangles), as a function of the inverse temperature. (c) Distribution of H-bond cluster sizes and (d) mean cluster size for each temperature.}
 \label{fig3}
\end{figure}
%%%%%%%%%%%%%%%%%%%%%%%%%%%%%%%%%%%%%%%%
Having pointed out the importance of cross-correlations in the dielectric response, let us examine how they build up spatially. To do so, we break them down as a function of the distance and the orientation with respect to a reference dipole as shown in fig.~\ref{fig2}a at 240~K. On average, molecules situated above and below the reference, even at a significant distance, are positively correlated, whereas lateral neighbors are negatively correlated. This corresponds to a favorable configuration with respect to dipole-dipole interaction. It is only at a very close distance that almost all neighboring molecules are positively correlated. What ultimately matters is the net correlation averaged over all neighboring molecules. If we restrict the sum to a cavity of radius $r$, this leads to the $r$-dependent Kirkwood correlation factor $G_\mathrm{K}(r)$:
\begin{equation}
 G_\mathrm{K}(r) = \frac{1}{\mu^2}\left\langle \vec{\mu}_i \cdot \sum_{r_{ij}<r} \vec{\mu}_j \right\rangle
 \label{eq_def_GKr}
\end{equation}
It is shown in fig.~\ref{fig2}b for each temperature. It displays a short-scale increase followed by a plateau, corresponding to a perfect cancellation of positive and negative cross-correlations. It is reached, independently of the temperature, after only two molecular distances, and its value gives the infinite-system size Kirkwood factor $g_\mathrm{K}$~(see Appendix B). Similarly to static correlations, dielectric spectra can be decomposed by considering virtual cavities of increasing radii, as shown in fig.~\ref{fig2}c for $T=220$~K. As we can see, cross-correlations gradually build up, with further dipoles leading to increasingly slower contributions. This observation is in agreement with solvation dynamics experiments on 1-propanol \cite{weigl2019local} showing that dipole relaxation at a local scale has a dynamic much closer to that of the self than the one obtained from DS. The final shape of the spectrum is obtained for a cavity including the fourth neighboring shell ($r_\mathrm{c}=20$~\AA). Interestingly, the closest neighbors taken alone show a contribution to the high-frequency flank of the spectral process that is not present anymore when all cross-correlations are considered, meaning that some high-frequency fluctuations are internally compensated. This is illustrated by the decomposition of the time correlation function into shells of increasing radius shown in fig.~\ref{fig2}d: the initial decorrelation of the first layer of neighbors (in green) is compensated by an increase in the correlation of further dipoles (in red). To further characterize the dynamics of cross-correlations, we show in fig.~\ref{fig2}e the relaxation time $\tau_\mathrm{tot}(r)$ in the cavity as a function of its radius. In contrast with the static cross-correlation $G_\mathrm{K}(r)$, which always reaches its final value after only 10~\AA, the relaxation time requires to include more distant neighbors to reach a plateau. The exact distance is hard to precisely estimate but close to 15~\AA~at 340~K and to 20~\AA~below 220~K. This spatial decomposition shows that the dynamic probed by DS is occurring over a radius of a few molecular distances, implying \textit{a priori} a significant number of molecules. Yet, it gives little information on the mechanism in itself and, in particular, on the role of HBs.

%%%%%%%%%%%%%%%%%%%%%%%%%%%%%%%%%%%%%%%%
\begin{figure*}[htbp]
 \centering
 \includegraphics[width=\linewidth]{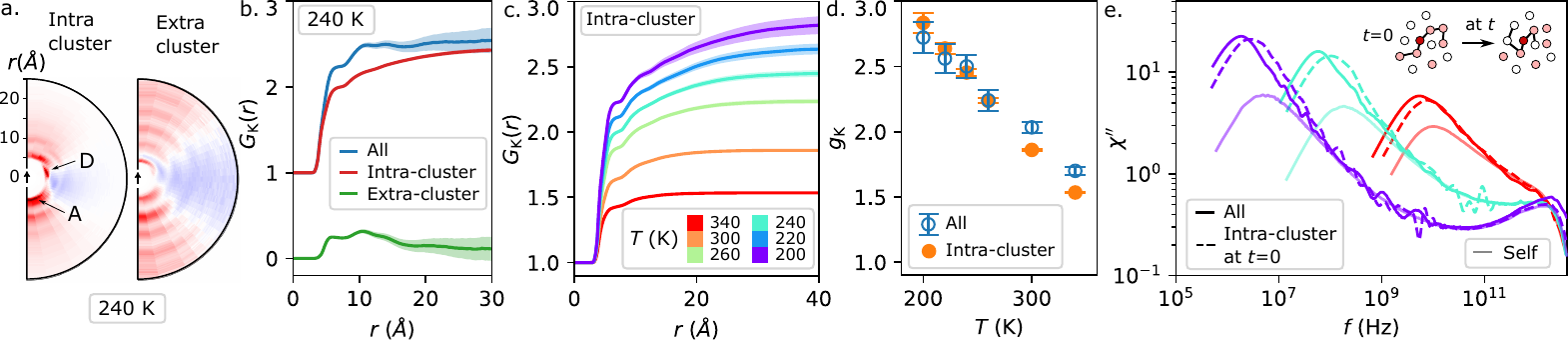}
 \caption{(a) Same as fig.~\ref{fig2}a, restricted to molecules within the same cluster as the reference molecule (left) or outside of it (right). (b) Decomposition of the $r$-dependent Kirkwood correlation factor $G_\mathrm{K}(r)$ at 240~K into intra and extra-cluster contributions. (c) Intra-cluster contribution to $G_\mathrm{K}(r)$ at different temperatures. (d) Finite-size Kirkwood correlation factor including all cross-correlations ($G_\mathrm{K}(r_\mathrm{c})$, in blue) and restricted to intra-cluster (in orange). (e) Same as fig.~\ref{fig1}a with the spectra (in dashed lines) at 340, 240, and 200~K, obtained by restricting the cross-correlations to the molecules within the same cluster as the reference molecule at $t=0$ (shown in red in the drawing).}
 \label{fig4}
\end{figure*}
%%%%%%%%%%%%%%%%%%%%%%%%%%%%%%%%%%%%%%%%
To move forward on this issue, let us characterize in detail the presence of HBs in the simulation. With a non-polarizable force field, HBs are mimicked through electrostatic interactions between fixed point charges. This simplification has proven acceptable for qualitatively reproducing their dynamics~\cite{laage2008molecular} and is required to reach fairly low temperatures with a reasonable system size. HBs between the hydroxyl group of a molecule (the donor) and the oxygen of another one (the acceptor) were identified using a geometrical criterion~\cite{luzar1996hydrogen, starr2000hydrogen, laage2006molecular, baptista2023chilling} (see Appendix C).  To characterize the dynamics of individual HBs, we define a correlation function $C_\mathrm{HB}(t)=\langle c_i(t)\rangle_i$, where $c_i(t) = 1$ only if donor $i$ has the same acceptor at times $t=0$ and $t$, else $0$. From its Fourier transform, a spectrum $\chi^{\prime\prime}_\mathrm{HB}$ can be obtained and is shown in fig.~\ref{fig3}a at several temperatures. It displays a peak, associated with a time $\tau_\mathrm{HB}$, at a frequency close to that of the self dipole orientational spectra $\chi^{\prime\prime}_{\mu~\mathrm{self}}$. The $\alpha$ process is thus clearly related to the dynamics of HBs: the full reorientation of a donor occurs when it changes acceptor, as previoulsy observed in water~\cite{starr2000hydrogen, laage2006molecular}. The fact that $\tau_\mathrm{HB}$ is much closer from $\tau_\alpha$ than $\tau_\mathrm{D}$ is consistent with experimental observations combining DS and NMR~\cite{gainaru2010nuclear} or infrared spectroscopy~\cite{gainaru2011hydrogen}. At high frequencies $\chi^{\prime\prime}_{\mu~\mathrm{self}}$ shows a more active dynamics than $\chi^{\prime\prime}_\mathrm{HB}$ related to changes in conformation affecting the dipole orientation without breaking HBs. The temperature evolution of the fraction $f$ of molecules involved in HBs is shown in fig.~\ref{fig3}b. At 340~K, 85~\% of molecules are involved either as a donor or acceptor, the rest being \textit{lone} molecules. Upon cooling, this fraction quickly increases and reaches 99~\% at 200~K. Interestingly, the temperature anomaly at 250~K reported in mono-alcohols~\cite{bauer2013debye} corresponds to the point at which lone molecules almost disappear. The fraction of molecules involved in HBs both as a donor and as an acceptor shows a significant temperature evolution over the whole accessible temperature range, going from 45 to 85~\%. Such large fractions are associated with the presence of supramolecular structures, denoted here as \textit{clusters}, which can be characterized from snapshots of the simulations as detailed in Appendix C. Because each molecule can accept as many as two donors, clusters can form branches and loops. Fig.~\ref{fig3}c shows, for each temperature, the distribution $P_n$ of cluster sizes, consistent with previous simulation results relying on a different force field \cite{baptista2023chilling}. Because the kinetics of HB formation and breaking cannot be described by constant rates~\cite{luzar1996hydrogen}, the exponential distribution characteristic of living polymer~\cite{douglas2006does} is not observed, especially at low temperature. The mean cluster size $\langle n P_n\rangle$ is shown as a function of temperature in fig.~\ref{fig3}d. It increases significantly upon cooling, going from 3 at 340~K to 18 at 200~K, in a comparable manner to what was reported on other mono-alcohols using near-infrared spectroscopy~\cite{bauer2013debye}. 

We now investigate the role of HB clusters in the static dielectric response. For this, we compute again the cross-correlation with respect to a reference molecule $i$: $\vec{\mu}_i\cdot\vec{\mu}_j$ but this time by separating the sum between molecules $j$ depending on whether or not they are part of the same cluster as $i$. This leads to the two diagrams shown in fig.~\ref{fig4}a whose sum would reproduce that of fig.~\ref{fig2}a. On the left one, corresponding to intra-cluster correlations only, the positive contributions of the direct acceptor and donors of $i$ are clearly visible (see arrows). Dipoles further away in the cluster are, on average, mostly positively correlated as well. Molecules outside of the cluster (see fig.~\ref{fig4}a, right) are undoubtedly also influenced by molecule $i$ to orient positively. This is well visible in fig.~\ref{fig4}b showing the $r$-depedent Kirkwood correlation factor and its decomposition into intra and extra clusters contributions. It appears that, while extra-cluster molecules play a role in how $G_\mathrm{K}$ builds up at short range, the dominant contribution originates mostly from within clusters. Fig.~\ref{fig4}b shows the intra-cluster contribution to $G_\mathrm{K}(r)$ at each temperature. The amplitude of the first step, corresponding to the first neighbor layer, is very temperature dependent, likely because of the evolution of the number of direct donors and acceptors. The second increase, related to molecules further away in the cluster, displays an even more dramatic temperature dependence, both regarding the amplitude and the scale over which it takes place, presumably due to the evolution of the mean size of clusters. The length scale suggests that there is a contribution arising from clusters in the upper part of the $P_n$ distribution, which, while being rare, involves a large number of molecules. When measuring the dielectric permittivity, the relevant quantity is the plateau value $g_\mathrm{K}$~\cite{henot2025}. Its temperature evolution is shown in fig.~\ref{fig4}d (in blue) together with the intra-cluster contribution (in orange). For $T \geq 300$~K, there is a net extra-cluster contribution, albeit small. However, this is no longer the case at lower temperatures. There, the net cross-correlation within the virtual cavity, which contains approximately 280 molecules, is, up to the resolution of the simulation, the same as that originating from the much fewer molecules which, on average, are part of the same HB cluster. 

Investigating the link between the dynamics of cross-correlations and HB clusters raises the question of cluster time evolution, which is particularly challenging to address for two main reasons. First, clusters identified from simulation snapshots are only partially stable, with only substructures persisting over timescales comparable to $\tau_\alpha$ (see Appendix D). Second, since $\tau_{\mathrm{HB}} \approx \tau_\alpha$, the clusters are intrinsically short-lived, with the breakup time of a cluster of size $N$ scaling as $\tau_\mathrm{HB}/N$~\cite{patil2023molecular}. This last point contradicts the assumption of the LPM ($\tau_\alpha \ll\tau_\mathrm{HB}$) that allows for sub-chains to relax while conserving their connectivity~\cite{patil2023molecular, cheng2024hydrogen}. To circumvent this difficulty, we adopted the simple approach illustrated by the drawing of fig.~\ref{fig4}d: for each molecule in the box, we considered the time correlation function originating only from the molecules that were part of the same cluster at time zero. This leads to the relaxation spectra shown as dashed lines in fig.~\ref{fig4}e for three temperatures. Their amplitudes are close to that of the total dielectric spectra, and their peak frequency lies between that of the $\alpha$ and of the Debye processes. This means that the initial orientation of the reference molecule can be retained, even after its relaxation, by molecules that were initially in the same cluster. Still, their decorrelation being faster than the Debye process indicates that the long-time dynamic results from molecules that were not part of the cluster at $t=0$. The spatial decomposition of dynamical cross-correlations of fig.~\ref{fig2}d provides information on the role of these molecules: those situated between 10 and 20~\AA~are only weakly correlated to $i$ at time zero, but this correlation increases and reaches a maximum after $\tau_\mathrm{self}$. Such behavior could be rationalized within the following picture: a given molecule exhibits a net positive orientational correlation with its neighbors, in which HB supramolecular structures play a key role in extending their spatial range and stabilizing them over time. As clusters evolve through bond breaking and the incorporation of new molecules, they favor the positive orientation of newcomers, even after the reference molecule itself has left the cluster. This mechanism allows the local dipole moment to persist on time scales longer than those of individual molecular reorientations. On average, this applies to any chosen reference molecule: the total cross-correlation is distributed over many molecules, with each contributing only a small fraction. This physical picture shares features with the DLO scenario~\cite{wieth2014dynamical}, in which HB chains polarize their environment, although the large clusters we observe encourage us to think that the spatial extension of the polarization arises through new molecules joining the cluster.

With a standard all-atom MD simulation, we resolved the experimental dielectric spectrum of 1-propanol between 340~K and 200~K, covering the significant increase in dielectric permittivity around the 250~K mono-alcohol anomaly~\cite{bauer2013debye}. By disentangling self and cross dipolar correlations, we showed that the $\alpha$ relaxation is related to the HB breaking dynamic, while the prominent Debye peak is the result of dipolar cross-correlations over a typical length-scale of 20~\AA. HB supramolecular clusters play a major role in the static dielectric response, although extra-cluster neighboring molecules also contribute to a lesser extent, especially above 300~K. Without having to characterize their time evolution, we show that HB clusters stabilize cross-correlations over durations larger than the relaxation time of the individual molecules, despite not preserving their connectivity over these durations. In the future, the methods developed here could be extended to other systems forming HB chains, rings or networks, providing deeper insight into how such supramolecular structures influence dielectric spectra. Moreover, the question remains of how the observations reported here, obtained at relatively high temperatures, evolve upon approaching the glass transition, where the separation between the $\alpha$ and Debye processes becomes much larger.

\begin{acknowledgments}
The authors thank T. Blochowicz, F. Pabst, R. Zei{\ss}ler, and F. Ladieu for fruitful discussions, and P. Lunkenheimer for kindly providing experimental dielectric spectra of 1-propanol. This project was provided with computing HPC and storage resources by GENCI at TGCC thanks to the grant 2025-16286 on the supercomputer Joliot Curie's ROME and V100 partition. 
\end{acknowledgments}

%\bibliography{biblio}% Produces the bibliography via BibTeX.


\onecolumngrid

\vspace{12pt}
\noindent\hrulefill \hspace{24pt} {\bf Appendix} \hspace{24pt} \hrulefill
\vspace{12pt}

\twocolumngrid

\renewcommand\thefigure{A\arabic{figure}} 
\renewcommand\thetable{A\arabic{table}} 

\renewcommand{\theequation}{A\arabic{equation}}
\setcounter{equation}{0}
\setcounter{figure}{0}

\textbf{A. Simulation details.} Simulations were performed using OpenMM~\cite{openmm_2017} with the OPLS-AA 3SSPP force-field~\cite{alva2022improving}, which is a reparametrization of OPLS-AA~\cite{jorgensen1996development} aimed at reproducing the dielectric constant and diffusion coefficient of 1-propanol at 300~K. The system consisted of $9600$ molecules in a cubic cell of side length $102$~\AA, with periodic boundary conditions. Bonds involving hydrogen were constrained, and the time step was 2~fs. The electrostatic interactions were computed using a Particle Mesh Ewald algorithm. At each temperature, an equilibration run was carried out in the NPT ensemble using a Monte Carlo barostat and a Nosé-Hoover thermostat during $t_\mathrm{eq}$ followed by a production run during $t_\mathrm{prod}$ in the NVT ensemble. Above 240~K, we used $t_\mathrm{eq}>300\tau_\mathrm{D}$ and $t_\mathrm{prod}>100\tau_\mathrm{D}$. At 220~K: $t_\mathrm{eq}=20\tau_\mathrm{D}$ and $t_\mathrm{prod}=70\tau_\mathrm{D}$, and at 200~K: $t_\mathrm{eq}=6\tau_\mathrm{D}$ and $t_\mathrm{prod}=45\tau_\mathrm{D}$. This last temperature accounted for 75~\% of the $2.6\times 10^9$ time steps run in this study.

\medskip

\textbf{B. Computation of relaxation spectra.} Relaxation spectra were obtained by considering the dipole moment of a virtual of radius $r_\mathrm{c} = 20~\AA$. The dipole correlation function of the cavity $C_\mu(t)$ was computed from equation~\ref{eq_def_Cmu} by averaging over all dipoles $i$ of the box and over initial times distributed over the
total duration of the simulation. As $(r_\mathrm{c}/a)^3 \approx 0.007 \ll 1$, this allows it to be almost unsensitive~\cite{henot2025} to the conductive electrostatic boundary condition induced by the use of the Ewald summation method. In other words, we get access directly to an estimate of the infinite size Kirkwood factor $g_\mathrm{K} \approx G_\mathrm{K}(r_\mathrm{c})$ rather than the finite size Kirkwood factor that would be obtained by considering the dipole moment of the whole simulation box. In this case, the frequency-dependent dielectric permittivity $\epsilon(\omega)$ is related to the correlation function of the cavity $C_\mu(t)$ by: 
\begin{equation}
 \frac{(\epsilon(\omega)-1)(2\epsilon(\omega)+1)}{\epsilon(\omega)} = -\frac {3\lambda}{\mu^2}\mathcal{L}_{i\omega}[\dot{C}_\mu(t)]
 \label{eq_epsw_GKt_rc1}
\end{equation}
Where $\mathcal{L}_{i\omega}$ is the one-sided Fourier transform, and $\lambda=\mu^2/(3\epsilon_0k_\mathrm{B}Tv)$ is a dimensionless parameter characterizing the strength of dipole-dipole interaction, with $v$ the molecular volume, $T$ the temperature, $k_\mathrm{B}$ the Boltzmann constant, $\epsilon_0$ the vacuum permittivity. Under the assumption that $\epsilon(\omega)\gg1$, it becomes:
\begin{equation}
 \epsilon(\omega)-1 = -\frac{3\lambda}{2\mu^2}\mathcal{L}_{i\omega}[\dot{C}_\mu(t)]
 \label{eq_epsw_GKt_rc2}
\end{equation}
This is the relation used in this work to obtain $\epsilon^{\prime\prime}(\omega) = -\mathrm{Im}(\epsilon(\omega)$. The Fourier transform of the correlation function was computed using the fftlog algorithm adapted to log-spaced data~\cite{Hamilton_2000}. The self part $\epsilon_\mathrm{self}(\omega)$ was obtained by replacing $C_\mu(t)$ by $C_\mathrm{self}(t)$ in eq.~\ref{eq_epsw_GKt_rc2}.

The relaxation spectra corresponding to the H-bonds dynamics were directly obtained from: $\chi_\mathrm{HB}^{\prime\prime} = -\mathrm{Im(\mathcal{L}_{i\omega}[\dot{C}_\mathrm{HB}(t)]))}$

\begin{figure}[htbp]
 \centering
 \includegraphics[width=\linewidth]{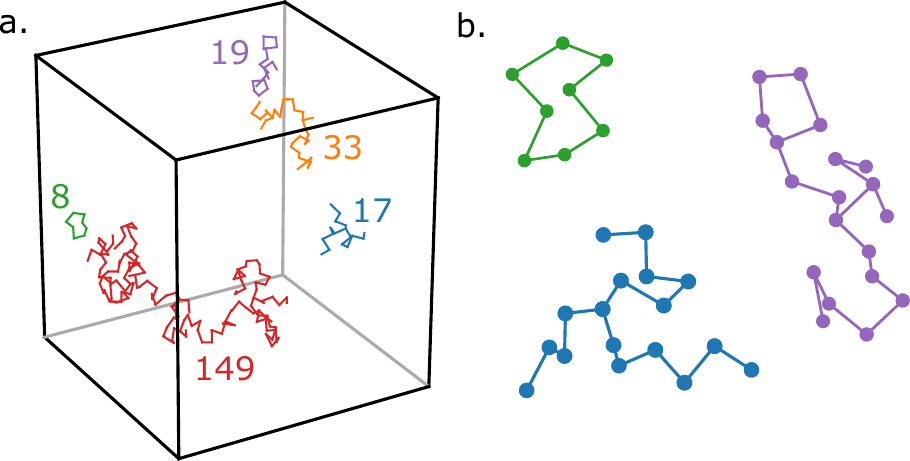}
 \caption{(a) Snapshot of the simulation box at 200~K showing five HB clusters and the number of molecules they include. (b) A close-up view of some clusters, showing loops and branches.}

 \label{suppl_cluster_repr}
\end{figure}

\medskip 
\textbf{C. Characterization of hydrogen bonds and clusters.}
Hydrogen bonds were identified using the MDAnalysis module~\cite{michaud2011mdanalysis} from the following geometrical criterion~\cite{luzar1996hydrogen, laage2006molecular, baptista2023chilling}: distance between donor and acceptor $\leq 3.5$~\AA~and O-H--O angle $>150$~°. We have checked that restricting the distance to 3.2~\AA ~does not significantly affect the results. HB clusters were reconstructed from the list of donor-acceptor at each timeframe using the algorithm given as supplemental material~\cite{SM} which takes into account the possibility of forming loops and branches. Typical clusters of various sizes are shown in fig.~\ref{suppl_cluster_repr}. In principle, percolating clusters, forming a loop across the periodic simulation box, could exist but we did not detect any occurrence at 200~K.

%%%%%%%%%%%%%%%%%%%%%%%%%%%%%%%%%%%%%%%%
%%%%%%%%%%%%%%%%%%%%%%%%%%%%%%%%%%%%%%%%

\begin{figure}[htbp]
 \centering
 \includegraphics[width=\linewidth]{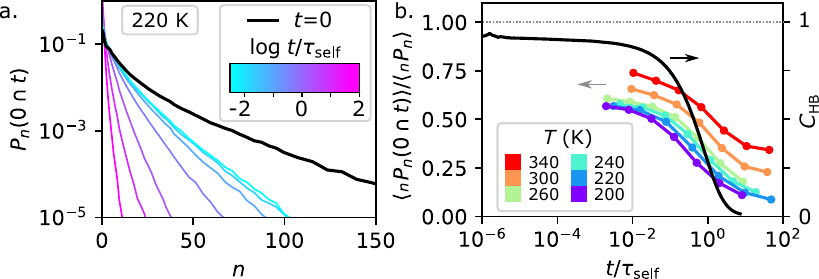}
 \caption{(a) Probability distribution $P_n(0\cap t)$ of finding $n$ molecules in common between a cluster at time zero and at time $t$. (b) For each temperature, the mean number of molecules in common normalized by the time zero value, as a function of the elapsed time normalized by the self relaxation time. The back curve indicates the individual HB correlation function at 200 K.}

 \label{suppl_cluster_dyn}
\end{figure}

\medskip 
\textbf{D. Dynamics of HB clusters.} While HB clusters can easily be characterized from snapshots of the simulation, following their time evolution turns out to be a challenging task. To illustrate this, let us consider the probability distribution of finding a certain number of molecules in common in a cluster between two snapshots separated by a time $t$. This is computed by taking a molecule $i$ as a reference, and considering the intersection between the clusters it is part of at times zero and $t$, before averaging over $i$. It is shown in fig.~\ref{suppl_cluster_dyn}a for 220~K for various times. Interestingly, the $t=0$ curve (in black) is very different from the one corresponding to very short times (less than $10^{-2}\tau_\mathrm{self}$). The time dependence of the mean number of molecules in common is shown in fig.~\ref{suppl_cluster_dyn}b for each temperature. The significant initial drop means that a fraction of the bounds are unstable and, even after very short times, clusters have already broken into sub-parts.
%%%%%%%%%%%%%%%%%%%%%%%%%%%%%%%%%%%%%%%%
%%%%%%%%%%%%%%%%%%%%%%%%%%%%%%%%%%%%%%%%
\end{document}